\documentstyle[preprint,aps,psfig,amsmath]{revtex}

\begin{document}

\draft

\title{Fluctuations in photon local delay time and their relation to phase spectra in random media}

\author{P. Sebbah$^1$, O. Legrand$^1$ and A.Z. Genack$^2$}
\address{$^1$Laboratoire de Physique de la Mati\`ere Condens\'ee/CNRS UMR 6622\\Universit\'e de Nice-Sophia Antipolis, Parc Valrose, 06108 Nice Cedex 02, France
\\$^2$Department of Physics, Queens College of the City University of New York, Flushing, New York 11367, USA}

\date{\today}

\maketitle

\begin{abstract}
The temporal evolution of microwave pulses transmitted through random dielectric samples is obtained from the Fourier transform of field spectra. 
Large fluctuations are found in the local or single channel delay time, which is the first temporal moment of the transmitted pulse at a point in the output speckle pattern. Both positive and negative values of local delay time are observed. The widest distribution is found at low intensity values near a phase singularity in the transmitted speckle pattern. In the limit of long duration, narrow-bandwidth incident pulses, the single channel delay time equals the spectral derivative of the phase of the transmitted field. Fluctuations of the phase of the transmitted field thus reflect the underlying statistics of dynamics in mesoscopic systems. 
\end{abstract}

\pacs{41.20.Jb, 05.40.+j, 71.55.Jv}

Statistical optics has concentrated on fluctuations of reflected and transmitted intensity. The Rayleigh distribution describes large fluctuations in intensity at the output of a scattering medium excited by monochromatic radiation. The intensity of a single polarization component normalized to its ensemble average has a negative exponential distribution, $\exp(-I/\left<I\right>) $\cite{Goodman}. This distribution obtains under the assumption that the field $E=Ae^{i\phi}$ can be represented as a superposition of uncorrelated partial waves. However, in multiply-scattering media, the coherent nature of wave propagation inevitably leads to both short and long-range intensity correlation \cite{Stephen,Feng,Genack}. These give rise to enhanced fluctuations in the intensity \cite{stretched}, total transmission \cite{Lagendijk,Kogan,Rossum,LangenBrouwerBeenakker,Marin}, and electronic conductance \cite{Laibowitz,LeeStone,Altshuler}, which have been studied intensively in the last decade. The degree of nonlocal intensity correlation is a measure of the closeness to the localization threshold \cite{Stephen,Feng,Genack,stretched,Lagendijk,Kogan,Rossum,LangenBrouwerBeenakker,Marin,Laibowitz,LeeStone,Altshuler} 
and determines the statistical distributions of key transmission quantities. In this paper we consider fluctuations in pulse propagation. This is in contrast to previous studies of the statistics of steady state propagation in random media and to measurements of the time of flight distribution\cite{Watson,GenackDrake,Alfano}. Fluctuations in the pulse evolution are averaged over an ensemble of samples and one obtains the arrival time distribution for transmitted photons, which is proportional to the path length distribution $P(s)$. This corresponds to the particle transport picture and gives a mix of ballistic and diffusive components. Here we consider fluctuations in the dynamics of transmission for a given incident and outgoing channel for different realizations of a random medium. Specifically, we define the local or single channel delay time, $\tau_{ab}$, as the first temporal moment for a transmitted pulse associated with an incident pulse of bandwidth $\Delta\omega$ centered at $t = 0$ in the time domain and at $\omega_0$ in the frequency domain,
\begin{equation}
\tau_{ab}(\omega_0,\Delta\omega)=
{\int|E_{ab}(t;\omega_0,\Delta\omega)|^2\,t\,\text{d}t
\over
\int|E_{ab}(t;\omega_0,\Delta\omega)|^2\,\text{d}t}\, , 
\label{taudef}
\end{equation}
where $E_{ab}(t;\omega_0,\Delta\omega)$ is the transmitted field in channel $b$ arising from an incident wave in channel $a$. This definition has been used in an earlier work in the context of nuclear physics \cite{Mello}. This is in contrast to previous discussions of  the Wigner time delay and the Wigner-Smith time delay matrix, which have been powerful concepts for a statistical description of scattering \cite{Wigner,Buttiker1,BrouwerFrahmBeenakker}. The diagonal elements $Q_{aa}$ of the lifetime matrix ${\bf Q}=-i{\bf S}^{-1}\partial{\bf S}/\partial\omega$ where ${\bf S}$ is the $2N\times 2N$ scattering matrix, are interpreted in terms of the time spent in the scattering region by a wave packet incident in one channel. As shown by Smith \cite{Smith}, they are the sum over all ouput channels (both in reflection and transmission) of the local delay time \cite{Eisenbud} weighted by the probability of emerging from that channel $I_{ab}/2N$: $Q_{aa}={1\over2N}\sum_b I_{ab}\tau_{ab}$. The sum of the $Q_{aa}$ over all $2N$ channels is the Wigner time delay $\tau_W=\sum_a Q_{aa}$ which is the trace of the lifetime matrix and is proportional to the density of states \cite{Iannaccone1,Buttiker2,Fyodorov}. Local delay times have been considered for electrons tunneling through barriers and for classical evanescent waves \cite{Landauer,Hauge}. In these cases, scattering through a fixed structure into a one dimensional system is considered. Here we study the propagation of an incident spatial mode into a multichannel random medium. The field is detected at a point in the output speckle pattern for an ensemble of random configurations. The subscript $ab$ which indicate the input and output channels is omitted in the following to simplify notation.

In a homogeneous medium of thickness $L$ with phase velocity $\nu$, the phase accumulated as the angular frequency is increased by $\Delta\omega$ is $\Delta\phi=\Delta kL=\Delta\omega L/\nu$, and is proportional to the delay time $\tau=L/\nu$ so that the delay time is $\tau=\Delta\phi/\Delta\omega$. In this simple situation, the phase derivative $\phi'\equiv\text{d}\phi/\text{d}\omega$ is known as the group delay and is a measure of the transit time through a homogeneous medium. In a random medium, however, the wave is multiply scattered and the output field is the superposition of partial waves arriving at a point. If we ignore fluctuations in the phase velocity for different paths in the medium and additional phase shifts associated with focal points and reflections, the phase accumulated by the wave in following a path of length $s$ as the frequency is incremented by $\Delta\omega$ is $\Delta\phi_s=\Delta\omega s/v$, giving a delay time $\Delta\phi_s/\Delta\omega=s/v$. The sum of these times weighted by $P(s)$, $\int P(s)s/v\text{d}s$, only gives the average value of the local delay time, $\langle\tau\rangle$. In this paper, we investigate the fluctuations of $\tau$. Measurements of microwave radiations propagating through random samples show that the single channel delay time varies with the width of the incident pulse. We find that fluctuations in $\tau$ increase as the pulse bandwidth narrows. For narrow bandwidth pulses, large positive and negative values of $\tau$ are found. These large values are related to the nature of the speckle pattern at the output of the medium. They occur most commonly when a null in the speckle pattern passes near the detector as the frequency is varied. The phase is undetermined at these nulls and jumps by $\pi$ radians when a null passes through the detector as the frequency is tuned. In contrast to the characteristics of static speckle patterns investigated by Freund which are independent of the scattering medium \cite{Freund}, here the phase variation reflects the underlying dynamics. When considering pulse propagation, it is natural to define the energy transmission coefficient $\epsilon_{ab}$ as well as the local delay time $\tau_{ab}$ of the ouput pulse. This gives a set of variables $(\epsilon,\tau)$. We will study pulses of specific width, particularly in the limit, $\Delta\omega\rightarrow0$. In this limit corresponding to long pulses, the local delay time $\tau$ approaches the spectral phase derivative $\phi'$ \cite{TOPS} and the energy transmission coefficient $\epsilon$ approaches the transmission coefficient $I$. 

The sample studied is composed of randomly positioned ${1\over 2}$-inch polystyrene spheres at a volume filling fraction of 0.52 contained within a one meter long, 7.6 cm diameter copper tube. New sample configurations are created by rotating the tube about its axis. Wire antennas are used as the emitter and detector at the input and output surface of the sample. A Hewlett Packard 8722C vector network analyzer performs a measurement of the microwave field, giving its amplitude and phase. Measurements are made between 7 and 25 GHz, using frequency intervals of 625 kHz. This wide frequency range allows us to reconstruct via Fourier transformation the time response to an input pulse of any given shape over a wide frequency range. The value of the local delay time depends upon the spectrum of the incident pulse. Here we consider incident pulses which are gaussian or rectangular in the frequency domain. The gaussian pulse has a rapid fall-off in both the time and frequency domains, whereas the rectangular pulse allows us to select a precise spectral range but oscillates in the time domain. Direct dynamical microwave measurements are possible in principle but generating precisely shaped pulses is not always practical. The complex response to an incident pulse with carrier frequency $\omega_0$ can be written as $E(t)=|E(t)|exp(\omega_0t+\phi(t))$. An example of the amplitude $|E(t)|$ and phase $\phi(t)$ of the response to a pulse constructed by Fourier transforming the field spectrum in a particular sample configuration is presented in Fig.~\ref{fig1} for two pulses centered at 10 GHz with a gaussian envelope $1/\sqrt{2\pi}\sigma\exp(-t^2/2\sigma^2)$ where $\sigma$ = 1 ns and 100 ns. 

Various properties have been used to characterize the travel time of a wavepacket \cite{Iannaccone2}. When the pulse is not appreciably distorted in transmission, the delay time of salient features such as the peak can be used. In multiply-scattering media, however, the shape is generally unrelated to that of the incident pulse and changes with configuration (see Fig.~\ref{fig1}a). It is therefore not possible to associate features of the transmitted pulse with the incident pulse. However, the shift of the barycenter of the transmitted pulse intensity at the output surface of the sample, as given is Eq.~(\ref{taudef}), is well defined even in a multiply-scattering system. We note that the integration over the pulse is reminiscent of the definition of the associated quantity, the energy transmission coefficient $\epsilon(\omega_0,\Delta\omega)=\int|E(t;\omega_0,\Delta\omega)|^2\text{d}t$, where the time origin is taken at the center of the incident pulse at the input surface. It is therefore natural to use Eq.~(\ref{taudef}) to represent the dynamic fluctuations of pulses in mesoscopic systems. We find $\tau$ = 41.8 ns and 59.8 ns for the pulses of Fig.~\ref{fig1}a and Fig.~\ref{fig1}b, respectively. The local delay time averaged over 581 sample configurations is $\langle\tau\rangle$ = 42.6 ns for a 1 ns incident pulse, while $\langle\tau\rangle$ = 45 ns for a 100 ns incident pulse. The difference in averaged values is due to variations in dynamical properties over the bandwidth. For comparison, the travel time through one meter of free space would be 3.3 ns. 

By changing the central frequency of the incident pulse, one can follow the variation with frequency of the single channel delay time for a given sample configuration and a given pulse bandwidth. The frequency dependence of $\tau(\omega,\Delta\omega)$ for a square pulse is plotted in Fig.~\ref{fig2} between 11 and 12 GHz for three different values of the incident pulse bandwidth $\Delta\omega$. A comparison with
\begin{equation}
{\Delta\phi\over\Delta\omega}(\omega) \equiv {\phi(\omega+\Delta\omega/2)-\phi(\omega-\Delta\omega/2)\over\Delta\omega}
\label{Deltaphi}\, ,\nonumber
\end{equation}
shows that $\tau$ and ${\Delta\phi}/{\Delta\omega}$ do not coincide as expected in such a medium. However, when the average of these quantities over sample configurations is taken, $\langle\tau\rangle$ is indistinguishable from $\langle{\Delta\phi}/{\Delta\omega}\rangle$ which is shown as the thin solid line. This figure also shows that fluctuations of $\tau$ and ${\Delta\phi}/{\Delta\omega}$ around their average values are of the same order of magnitude and increase with decreasing pulse bandwidth. The local delay time $\tau$ can be smaller than the travel time in free space and even negative. This occurs most frequently for pulses of duration greater than $\langle\tau\rangle$ corresponding to $\Delta\omega$ less than the field correlation frequency, $\delta\omega\sim\langle\tau\rangle^{-1}$. Fig.~\ref{fig3} shows the response to a gaussian pulse with $\sigma$ = 100 ns, centered at 11.1025 GHz, which corresponds to the first negative peak in Fig.~\ref{fig2}c.

In order to clarify the character of these fluctuations, we plot in Fig.~\ref{fig4} the phase derivative of the transmitted field, its phase modulus $2\pi$, and the logarithm of the transmitted intensity which represents the complete field between 10 and 10.5 GHz. Large positive and negative peaks in the phase derivative are associated with small values of intensity. Indeed a zero in the amplitude would correspond to an undefined phase since the real and imaginary part of the field are then zero. The equiphase line map around such a phase singularity is a ``star" \cite{Freund} and the phase circulation around this singularity is an integer multiple of $2\pi$ \cite{Berry}. A phase singularity is by convention positive if the phase circulates counterclockwise. Connection between equiphase lines of different phase singularities have been investigated by Freund \cite{Freund}. To explore excursions in the phase as a singularity moves near a point, we measured the phase at closely spaced points along a line as the frequency is tuned. After the spectrum at a given position is taken, the detector is translated by $\Delta x$ = 1 mm on a 4 cm-length line running symmetrically about the center of the output surface. The increment in phase along the line from 18 GHz is obtained by unwrapping the phase modulus $2\pi$ \cite{PRE}, which is shown in Fig.~\ref{fig5}a. As the speckle pattern changes with increasing frequency, phase singularities may move across the detection line resulting in a $2\pi$ phase difference between consecutive detector positions giving a $+\pi$ jump on one side of the singularity and $-\pi$ jump on the other side. The phase difference between two consecutive positions of the detector is presented in Fig.~\ref{fig6}. A $2\pi$ step is the signature of a phase singularity moving between these two positions. As $\Delta x$ goes to zero, this phase difference plot would become a series of sharp steps and flat plateaus. The presence of phase singularities results in large fluctuations of the phase derivative $\phi'$ as a function of frequency and of detector position as shown in Fig.~\ref{fig5}b. Strong fluctuations of $\Delta\phi/\Delta\omega$ will occur when $\Delta\omega$ is small since it is then not so different from $\phi'$ (Fig.~\ref{fig2}c). For larger $\Delta\omega$ (Fig.~\ref{fig2}a), averaging of random variations in phase reduces these fluctuations.

The magnitude of fluctuations of $\Delta\phi/\Delta\omega$ and $\tau$ are comparable as are their correlation frequencies. In the limit $\Delta\omega\rightarrow0$, these become identical. In Appendix A, we show that the delay time can be expressed in terms of the intensity and the phase derivative as,
\begin{equation}
\tau(\omega_0,\Delta\omega)= 
{\int g^2_{\omega_0,\Delta\omega}(\omega)I(\omega)\phi'\,\text{d}\omega
\over
 \int g^2_{\omega_0,\Delta\omega}(\omega)I(\omega)\,\text{d}\omega}\, ,
\label{Itau}
\end{equation}
where $g_{\omega_0,\Delta\omega}(\omega)$ is the spectrum of the incident pulse. For one particular sample configuration, the rhs of Eq.~(\ref{Itau}) is shown in Fig.~\ref{fig7} for a gaussian incident pulse with bandwidth $\Delta\omega = 1/2\pi\sigma$ = 79.6 MHz and central frequency $\omega_0$ between 7 and 25 GHz and is indistinguishable from the delay time $\tau$ of a gaussian pulse with $\sigma$ = 2 ns which is also plotted in Fig.~\ref{fig7}. Eq.~(\ref{Itau}) is found to be accurate experimentally to within 0.1 \% (1\%) for a gaussian pulse with $\sigma$ = 1 ns ($\sigma$ = 100 ns). Equation~(\ref{Itau}) provides a useful shortcut to the computation of the local delay time as compared to computing  the temporal integral of Eq.~(\ref{taudef}) from the Fourier transform of the field.

For pulse bandwidths much smaller than the correlation frequency, which is essentially the inverse of the average delay time, the transmission coefficient $I$ is roughly constant over the pulse bandwidth. Therefore, $\tau\sim \Delta\phi/\Delta\omega$ for a narrow rectangular pulse in the frequency domain (Fig.~\ref{fig2}c). Only at frequencies at which the intensity drops rapidly and cannot be taken as constant over the bandwidth are fluctuations in $\Delta\phi/\Delta\omega$ larger than fluctuations in $\tau$. In these cases, a phase singularity moves near the detector giving a change in phase of the order of $\pi$ over a frequency change $\Delta\omega$ that can be arbitrarily small and in particular smaller than $\delta\omega$. The shape of a pulse, which encompasses a bandwidth over which the intensity changes appreciably, will generally be different from
that of the incident pulse as is seen in Fig.~\ref{fig3}.

Taking the limit $\Delta\omega\rightarrow0$ in both Eqs.~(\ref{A4}) and (\ref{A5}) of Appendix A, one finds,
\begin{equation}
\underset{\Delta\omega\rightarrow0}{lim}\tau(\omega_0,\Delta\omega)=\phi'(\omega_0)\, .
\label{limit}
\end{equation}
To illustrate this result, the time delay between 10 and 11 GHz is presented in Fig.~\ref{fig8} for various pulse widths and is compared to the phase derivative $\phi'$ for the same frequency range (Fig.~\ref{fig8}e). We find experimentally that for a pulse with standard deviation $\sigma$ = 400 ns, which corresponds to a bandwidth of $\Delta\omega=1/2\pi\sigma \sim$ 0.4 MHz, the time delay and the phase derivative are almost indistinguishable.
Finally, considering the total energy of the output pulse $\epsilon(\omega_0,\Delta\omega)=\int|E(t;\omega_0,\Delta\omega)|^2dt$, we find from Parseval's theorem \cite{Goodman} that 
$\underset{\Delta\omega\rightarrow0}{\lim}\epsilon=I$.
We are able to define a general statistical set of variables in the time domain $(\epsilon,\tau)$ which capture the statistics of dynamics in mesoscopic systems and which approach the variables $(I,\phi')$ in the limit of narrow bandwidth pulses.

In conclusion, we have investigated the dynamics of wave propagation through random media by considering the energy transmission coefficient and the local delay time $(\epsilon,\tau)$, which, in the limit of long bandwidth limited pulses, approach the transmitted intensity and phase derivative, respectively $(I,\phi')$. We demonstrate that $\tau$ is the integral over frequency of $I\phi'$ weighted by the spectral density of the incident pulse. Fluctuations in $\tau$ are found to be particularly large, for narrow bandwidth pulses, when $\epsilon$ is small. This generally occurs when a phase singularity in the static speckle pattern passes near the detector and indicates that the statistics of $\tau$ and $I$ are related. Measurements \cite{PaperB} and calculations \cite{PaperC} of key dynamical distributions and correlation functions, which will be presented elsewhere, confirm the interplay between $I$ and $\phi'$. Here we have dealt with single channel quantities. When considering the spatially averaged delay time, however, it is appropriate to weight the local delay time by the energy transmission coefficient, which is $I\phi'$ in the limit of narrow bandwidth pulses. The sum of $I\phi'$ over all incident and outgoing channels is proportional to the density of states and is in many respects analogous to the conductance, which is the sum of $I$ over all input and output channels. We expect that large fluctuations in the spatially averaged delay time will occur as a result of spatial
correlation in $I\phi'$ just as enhanced conductance and transmission
fluctuations arise as a consequence of spatial correlation of $I$. Thus the
observations in this paper should form the basis for treating the dynamical
aspects of mesoscopic physics.

\acknowledgments
We thank Bart van Tiggelen, Fabrice Mortessagne and Marin Stoytchev for useful discussions. We are indebted to Narciso Garcia for valuable suggestions, for support and encouragement. This work was supported by the Groupement de Recherches POAN 1180, the National Science Foundation under Grant Nos. DMR 9632789 and INT9512975, and a PSC-CUNY grant.

\appendix
\section{}
To prove Eq.~(\ref{Itau}), we first express the transmitted intensity as
\begin{equation}
|E(t;\omega_{0},\Delta\omega)|^2=
\int\int\text{d}\omega_1\text{d}\omega_2\,
\tilde E^*(\omega_1-\omega_{0},\Delta\omega)
\tilde E(\omega_2-\omega_{0},\Delta\omega)
~e^{i(\omega_1-\omega_2)t}\, ,
\label{A1}
\end{equation}
where $\tilde E(\omega-\omega_{0},\Delta\omega)=
g_{\omega_0,\Delta\omega}(\omega)E(\omega)$
 is the Fourier transform of the transmitted field for a given pulse shape $g_{\omega_0,\Delta\omega}(\omega)$. Taking $\omega_1=\omega$ and $\omega_2=\omega+\eta$, we calculate the numerator of $\tau$ in Eq.~(\ref{taudef}):
\begin{eqnarray}
&&\int|E(t;\omega_{0},\Delta\omega)|^2t\,\text{d}t \nonumber\\
&&=\int\text{d}t
\int\text{d}\eta
\int\text{d}\omega\,
te^{-i\eta t}
\tilde E^*(\omega-\omega_{0},\Delta\omega)
\tilde E(\omega+\eta-\omega_{0},\Delta\omega) \nonumber\\
&&=\int\text{d}t
\int\text{d}\eta
\int\text{d}\omega\,
[i{\partial\over\partial\eta}(e^{-i\eta t})]
\tilde E^*(\omega-\omega_{0},\Delta\omega)
\tilde E(\omega+\eta-\omega_{0},\Delta\omega)\, .
\label{A2}
\end{eqnarray}
Assuming that $g_{\omega_0,\Delta\omega}(\omega)$ is vanishing at infinity, we find after integrating by parts,
\begin{eqnarray}
&&\int|E(t;\omega_{0},\Delta\omega)|^2t\,\text{d}t \nonumber\\
&&=-i\int\text{d}\eta
\int\text{d}t\,
e^{-i\eta t}
\int\text{d}\omega
\tilde E^*(\omega-\omega_{0},\Delta\omega)
{\partial\over\partial\eta}
\tilde E(\omega+\eta-\omega_{0},\Delta\omega) \nonumber\\
&&=-i\int\text{d}\eta\,
2\pi\delta(\eta)
\int\text{d}\omega\,
\tilde E^*(\omega-\omega_{0},\Delta\omega)
{\partial\over\partial\omega}
\tilde E(\omega+\eta-\omega_{0},\Delta\omega) \nonumber\\
&&=-2i\pi\int\text{d}\omega\,
\tilde E^*(\omega-\omega_{0},\Delta\omega)
{\partial\over\partial\omega}
\tilde E(\omega+\eta-\omega_{0},\Delta\omega)\, .
\label{A3}
\end{eqnarray}
Writing $\tilde E=|\tilde E|e^{i\phi}$, we obtain
\begin{equation}
\int|E(t;\omega_{0},\Delta\omega)|^2t\,\text{d}t
=2\pi\int|\tilde E(\omega-\omega_{0},\Delta\omega)|^2
\phi'(\omega)\,\text{d}\omega\, .
\label{A4}
\end{equation}
The calculation of the denominator of Eq.~(\ref{taudef}) is straightforward:
\begin{equation}
\int|E(t;\omega_{0},\Delta\omega)|^2\,\text{d}t
=2\pi\int|\tilde E(\omega-\omega_{0},\Delta\omega)|^2\,
\text{d}\omega\, ,
\label{A5}
\end{equation}
which gives Eq.~(\ref{Itau}).

\begin{figure}
\vspace{1.cm}
\caption{Amplitude of the time response (solid line) to a gaussian pulse (dashed line) with (a) $\sigma$ = 1 ns and (b) $\sigma$ = 100 ns, centered around 10 GHz. Total energy of the input pulse, $\int|E|^2\,\text{d}t$ is normalized to unity. The solid curve is the actual scale, while the dashed curve is rescaled. The slowly varying component of the phase $\phi(t)$ is shown as the dotted curve.}
\label{fig1}
\end{figure}

\begin{figure}
\vspace{1.cm}
\caption{Local delay time (thick solid line) for square bandwidth pulse with bandwidths of (a) $\Delta\omega$ = 500MHz, (b) $\Delta\omega$ = 50 MHz and (c) $\Delta\omega$ = 5 MHz compared to $\Delta\phi/\Delta\omega$ (dotted line) versus frequency. Averaged local delay time over 581 sample configurations is shown in the thin solid line.}
\label{fig2}
\end{figure}

\begin{figure}
\vspace{1.cm}
\caption{Amplitude time response (solid line) to a gaussian pulse (dashed line) with $\sigma$ = 100 ns, centered around 11.1025 GHz. Solid curve is the actual scale while dashed the curve is rescaled. The slowly varying component of the phase $\phi(t)$ is shown as the dotted curve.}
\label{fig3}
\end{figure}

\begin{figure}
\vspace{1.cm}
\caption{(a) Phase derivative, (b) phase modulus $2\pi$ and (c) logarithm of the magnitude between 10 and 10.5 GHz.}
\label{fig4}
\end{figure}

\begin{figure}
\vspace{1.cm}
\caption{Surface plot of (a) cumulative phase and (b) phase derivative with frequency between 18 and 18.1 GHz across  the output of the sample obtained by sweeping the detector over 40 mm, each millimeter. }
\label{fig5}
\end{figure}

\begin{figure}
\vspace{1.cm}
\caption{Phase difference between 4 consecutive detector positions ($x$ = 19 to 22 mm) at the output surface of the sample of Fig.~\ref{fig5} in the frequency range 18 to 19 GHz.}
\label{fig6}
\end{figure}

\begin{figure}
\vspace{1.cm}
\caption{Right hand side of Eq.~(\ref{Itau}) between 7 and 25 GHz for one sample configuration when considering a gaussian pulse shape of 79.6 MHz bandwidth. This plot is indistinguishable from the plot of the delay time versus frequency of a gaussian pulse with $\sigma$ = 2 ns.}
\label{fig7}
\end{figure}

\begin{figure}
\vspace{1.cm}
\caption{Time delay between 10 and 11 GHz  for an input pulse with (a) $\sigma$ = 10 ns, (b) $\sigma$ = 30 ns, (c) $\sigma$ = 100 ns, (d) $\sigma$ = 300 ns. The phase derivative for the same frequency range is shown in (e).}
\label{fig8}
\end{figure}

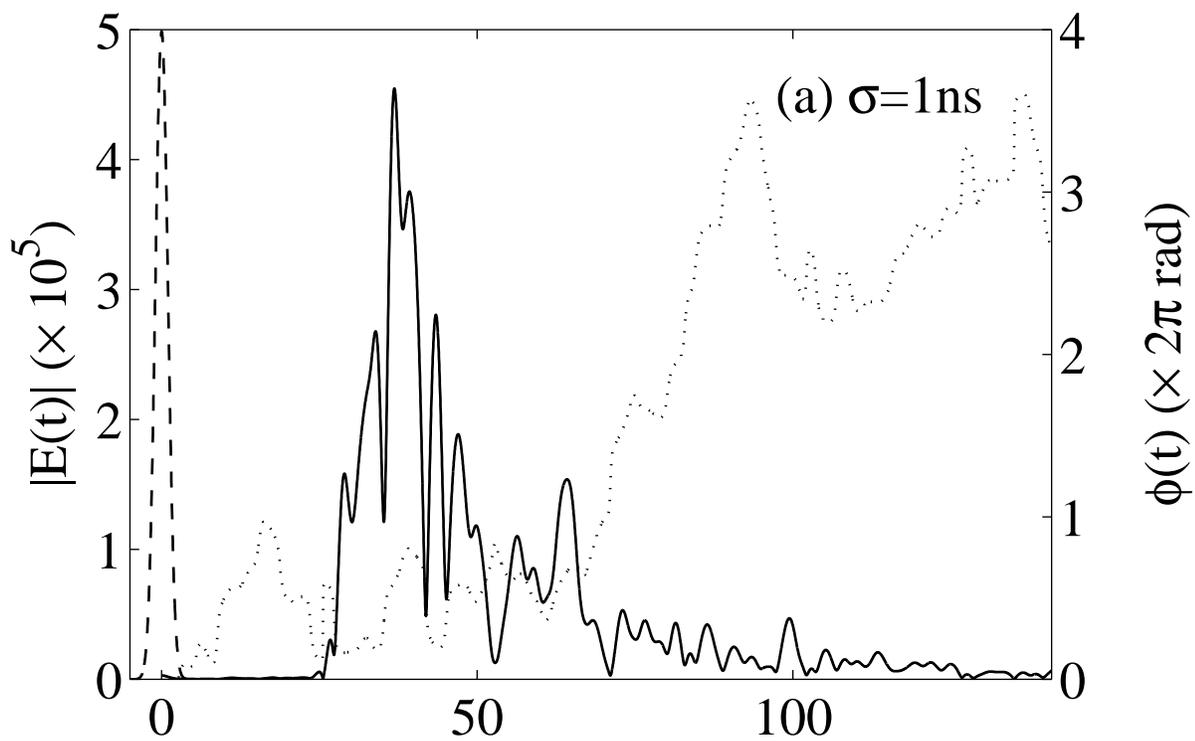
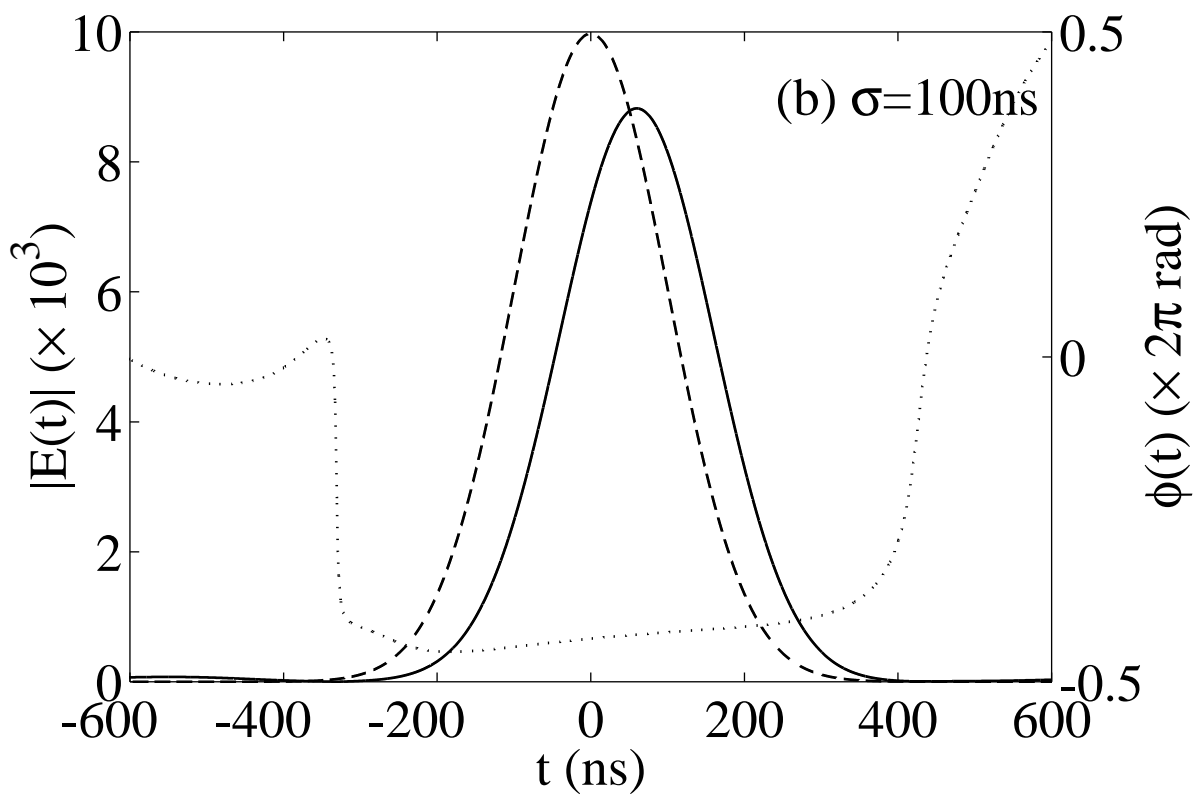
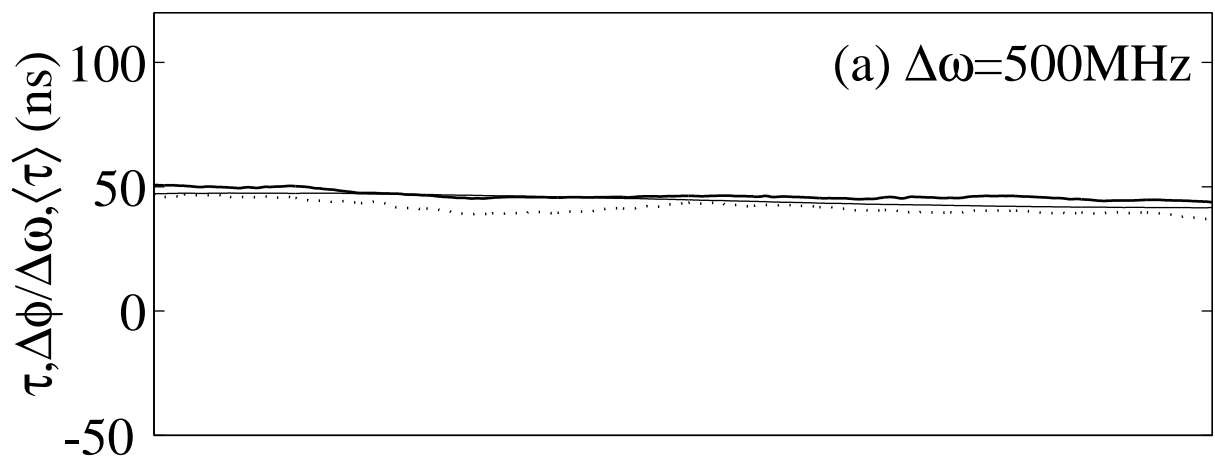
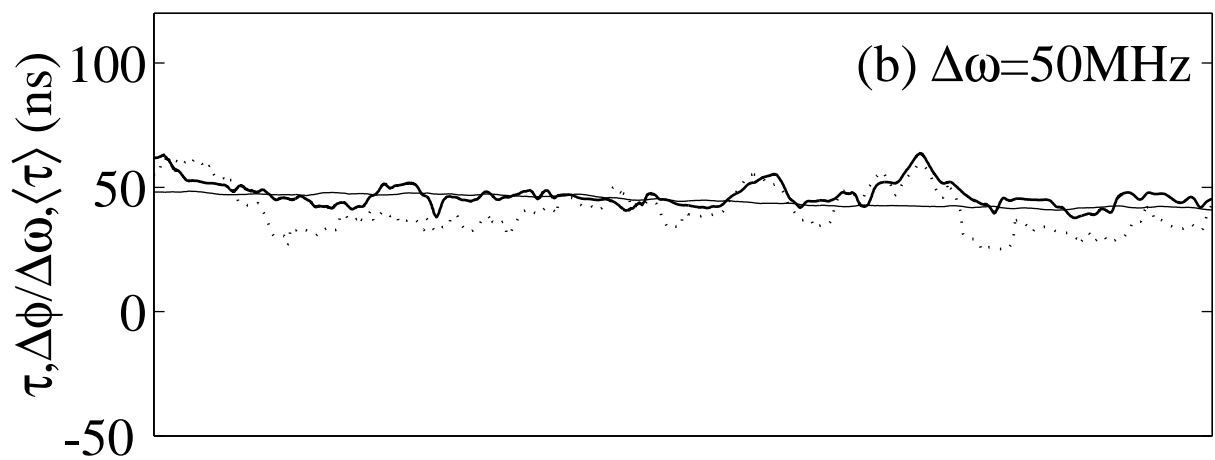
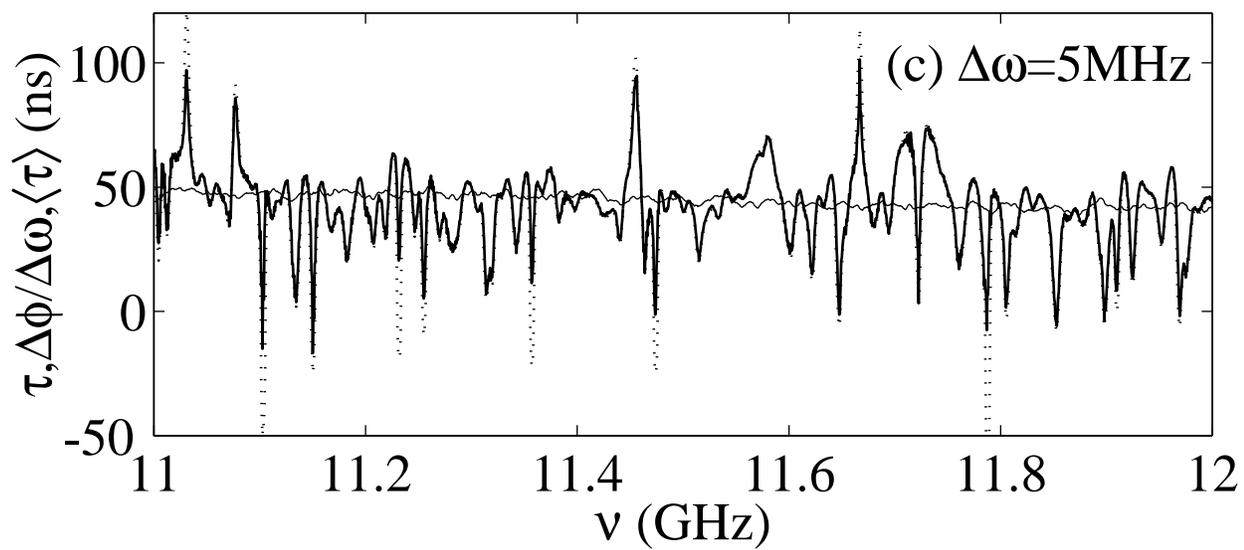
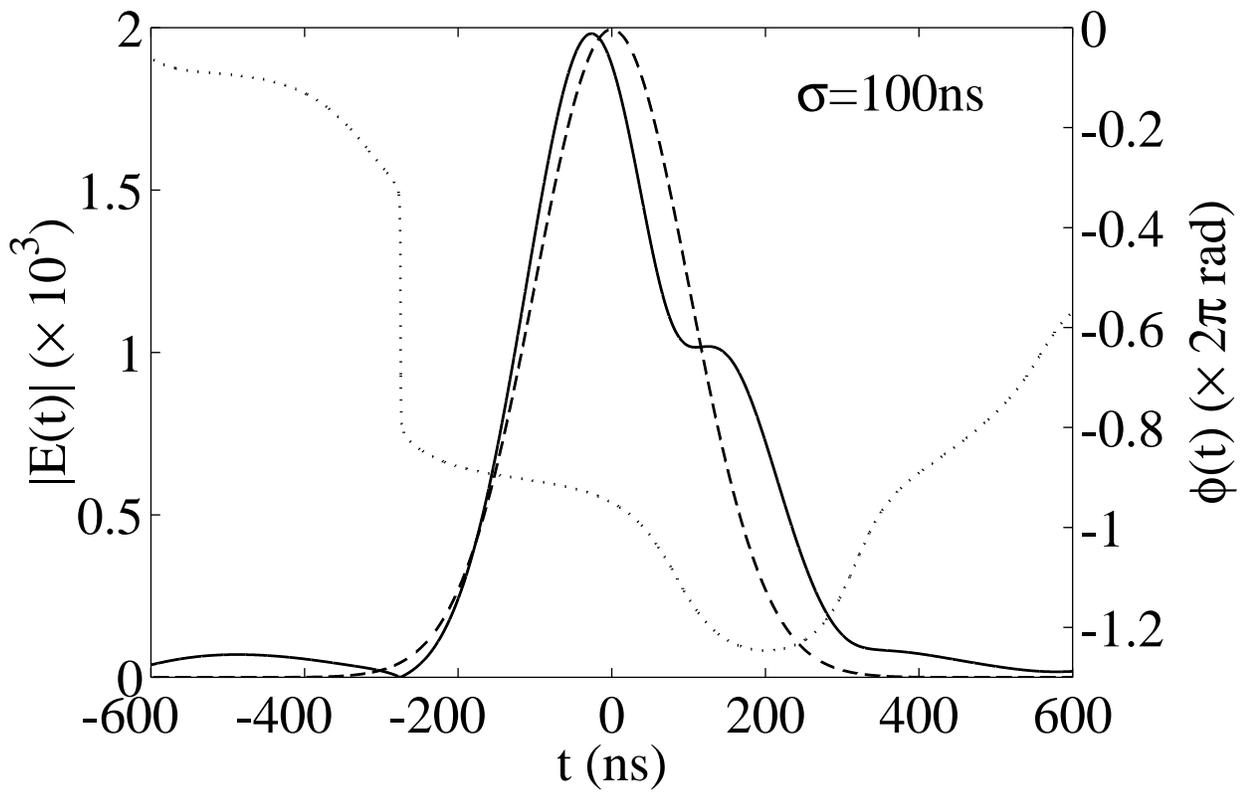
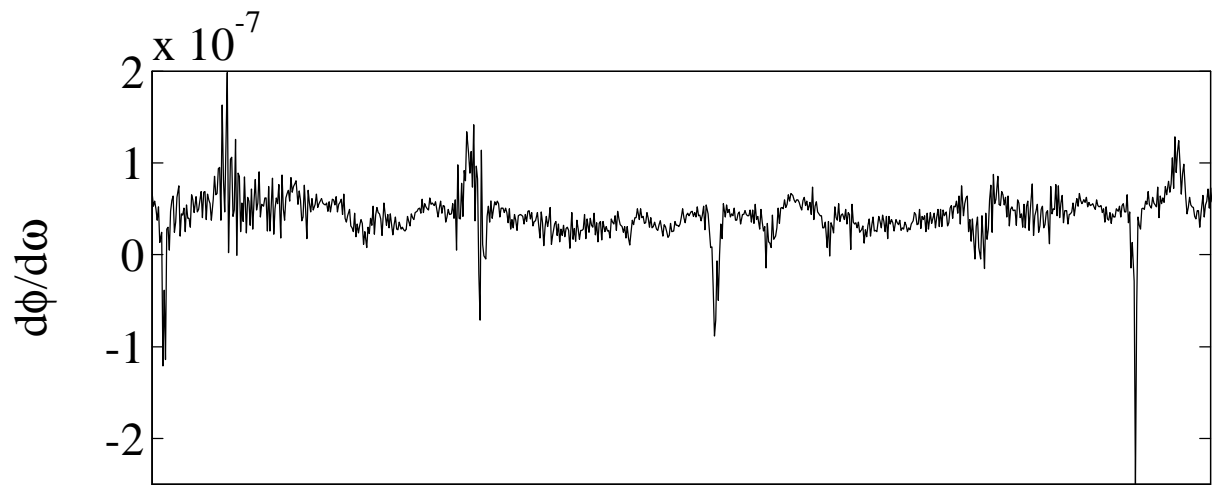
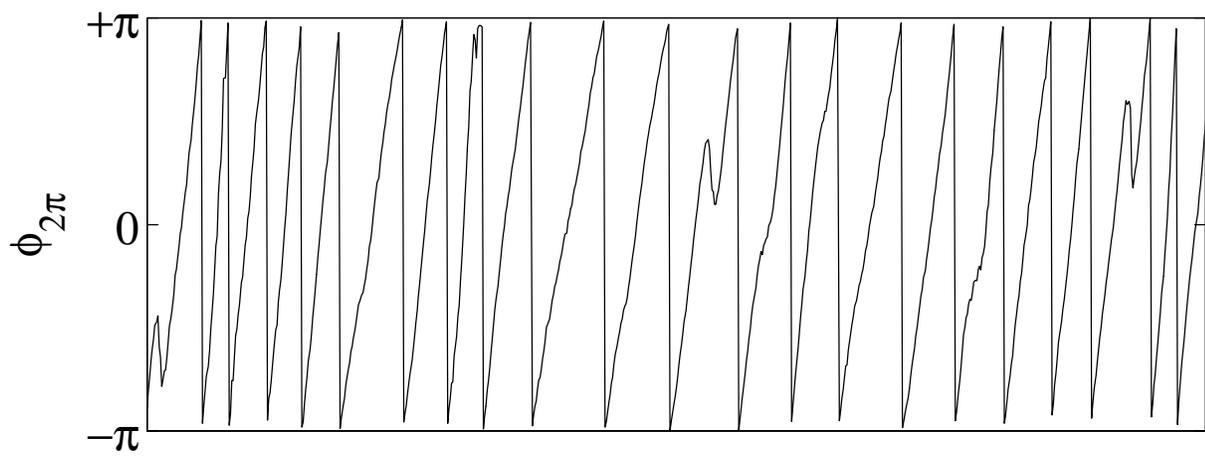
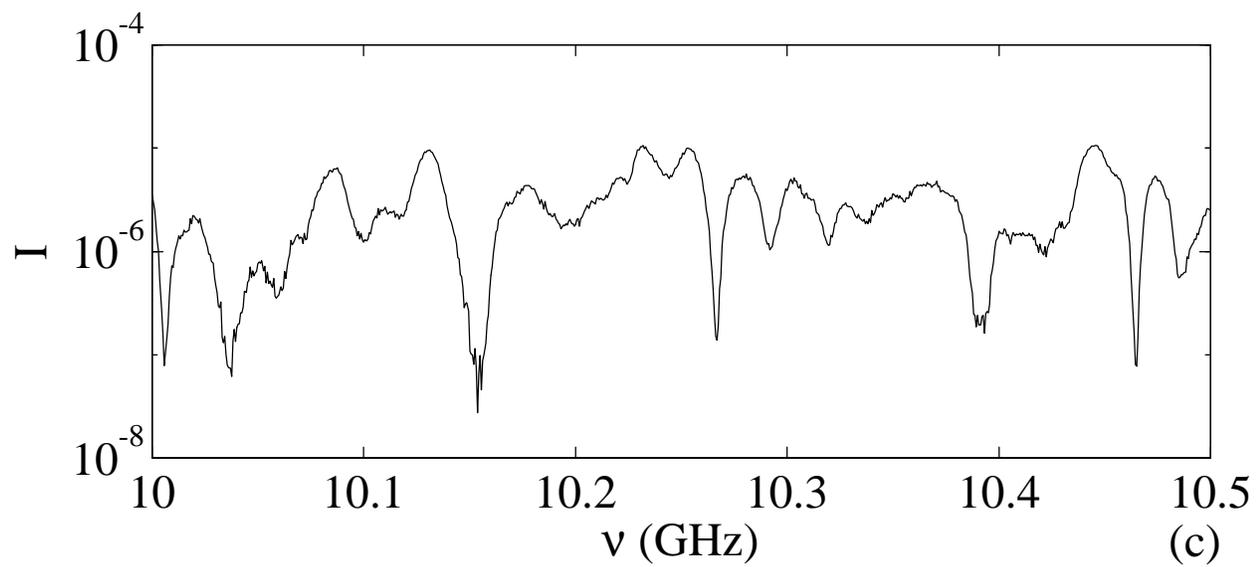
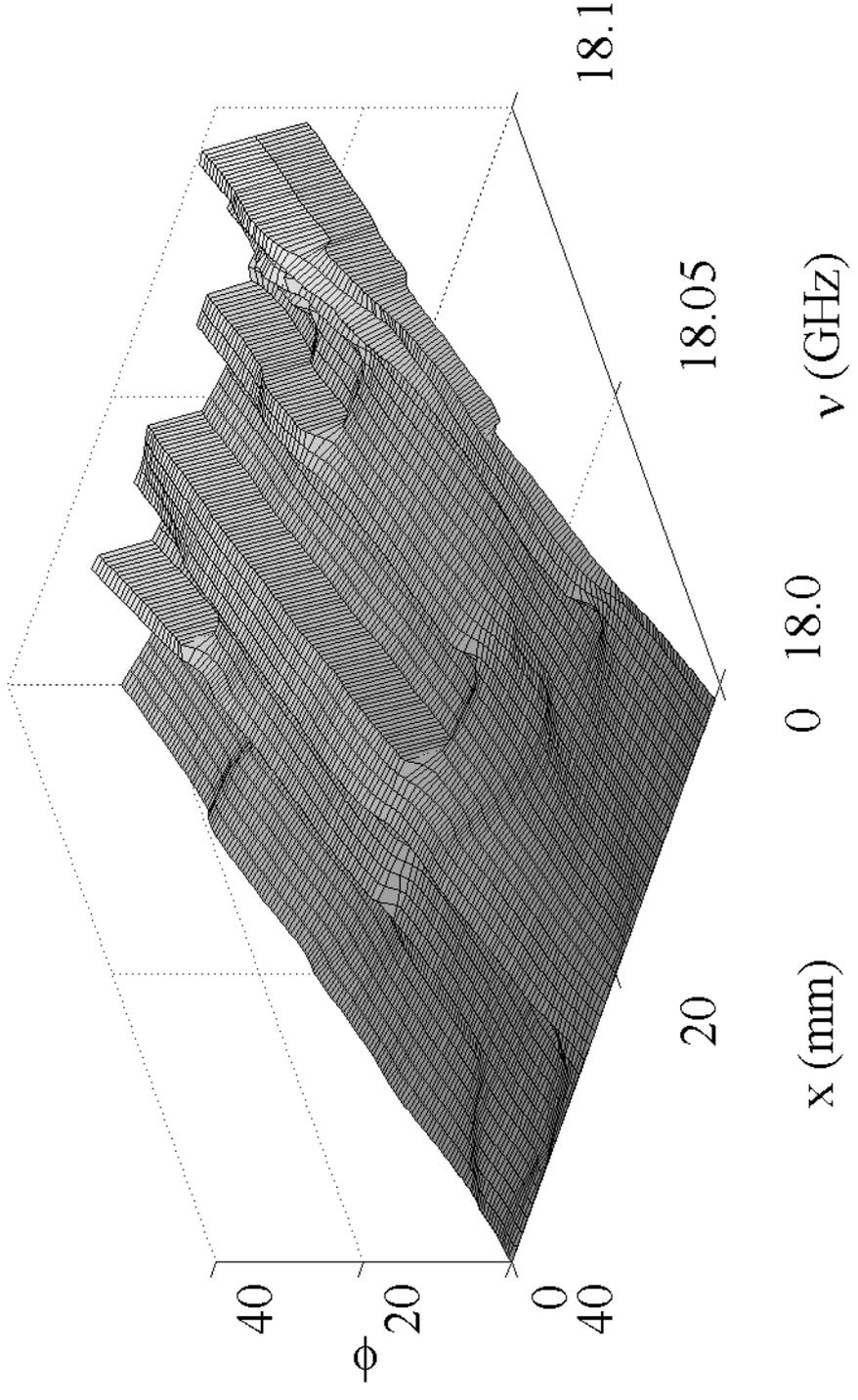
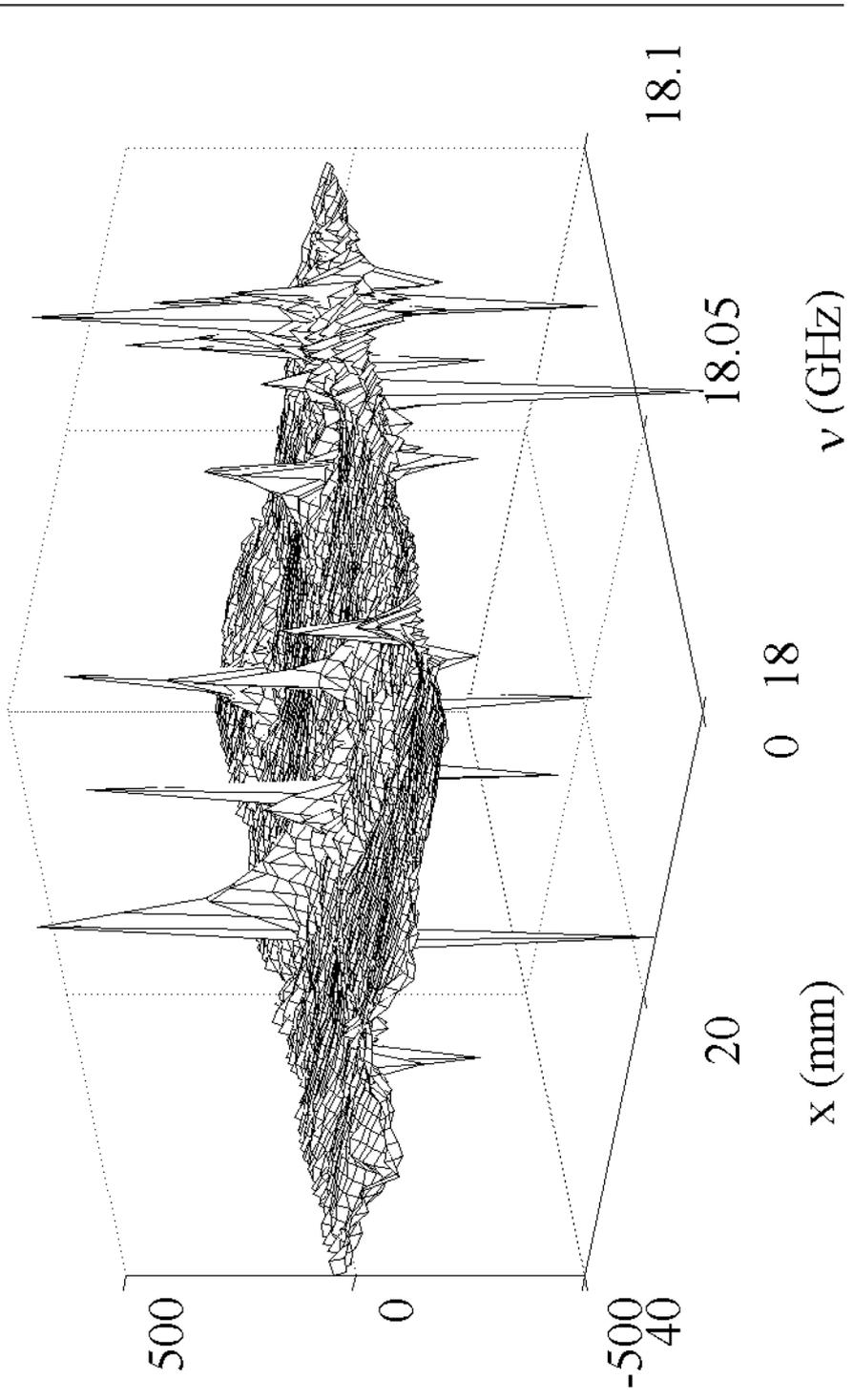
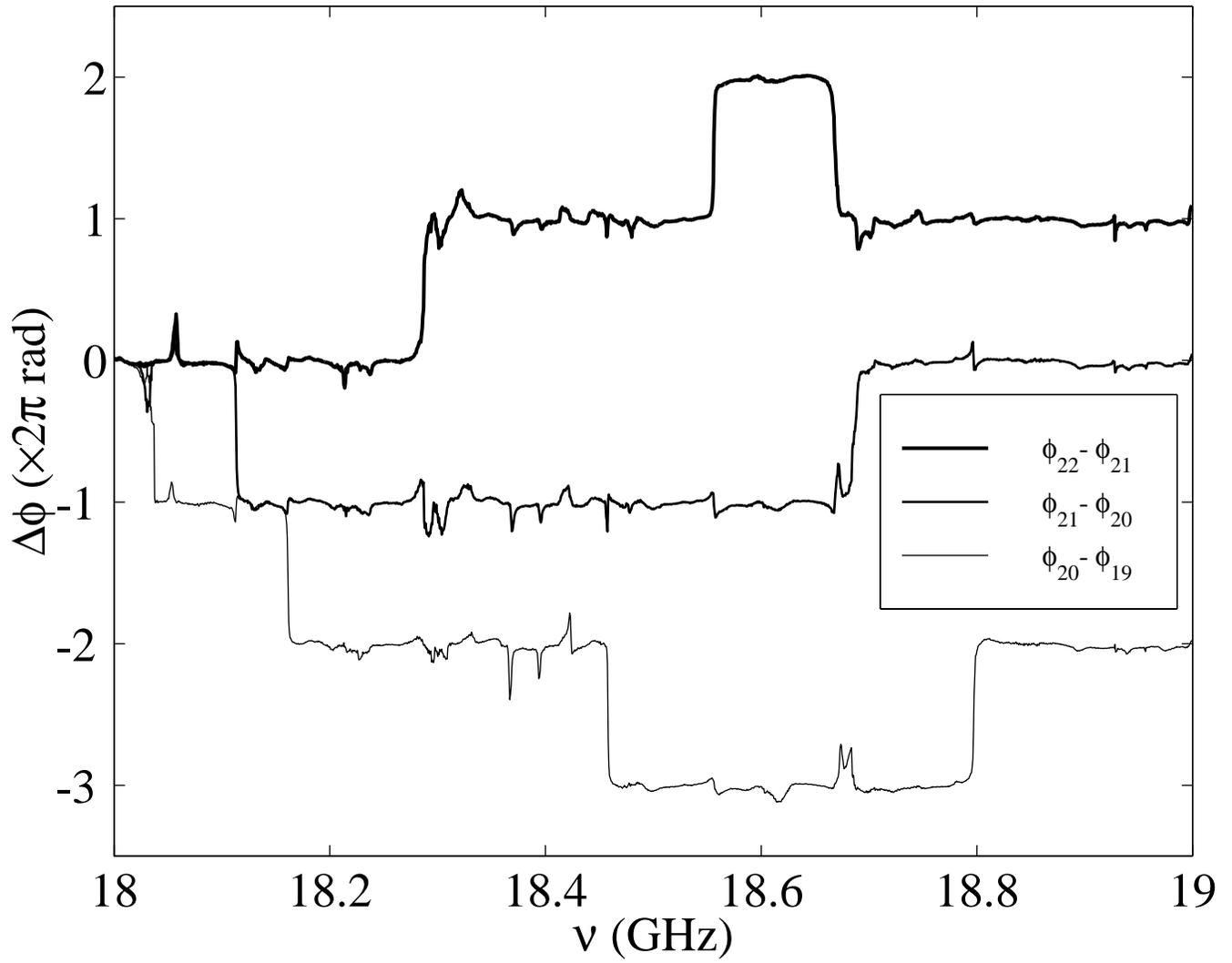
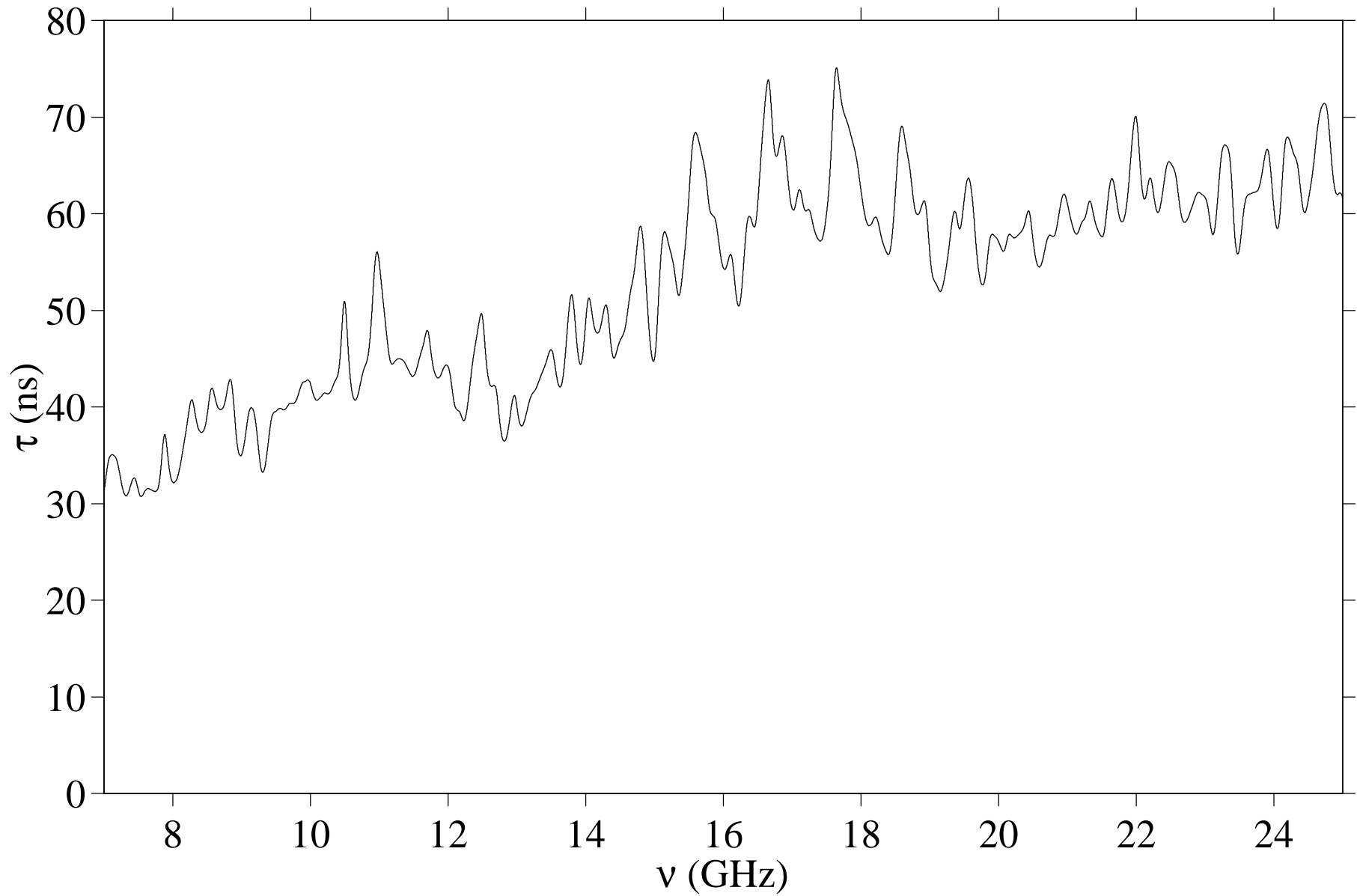
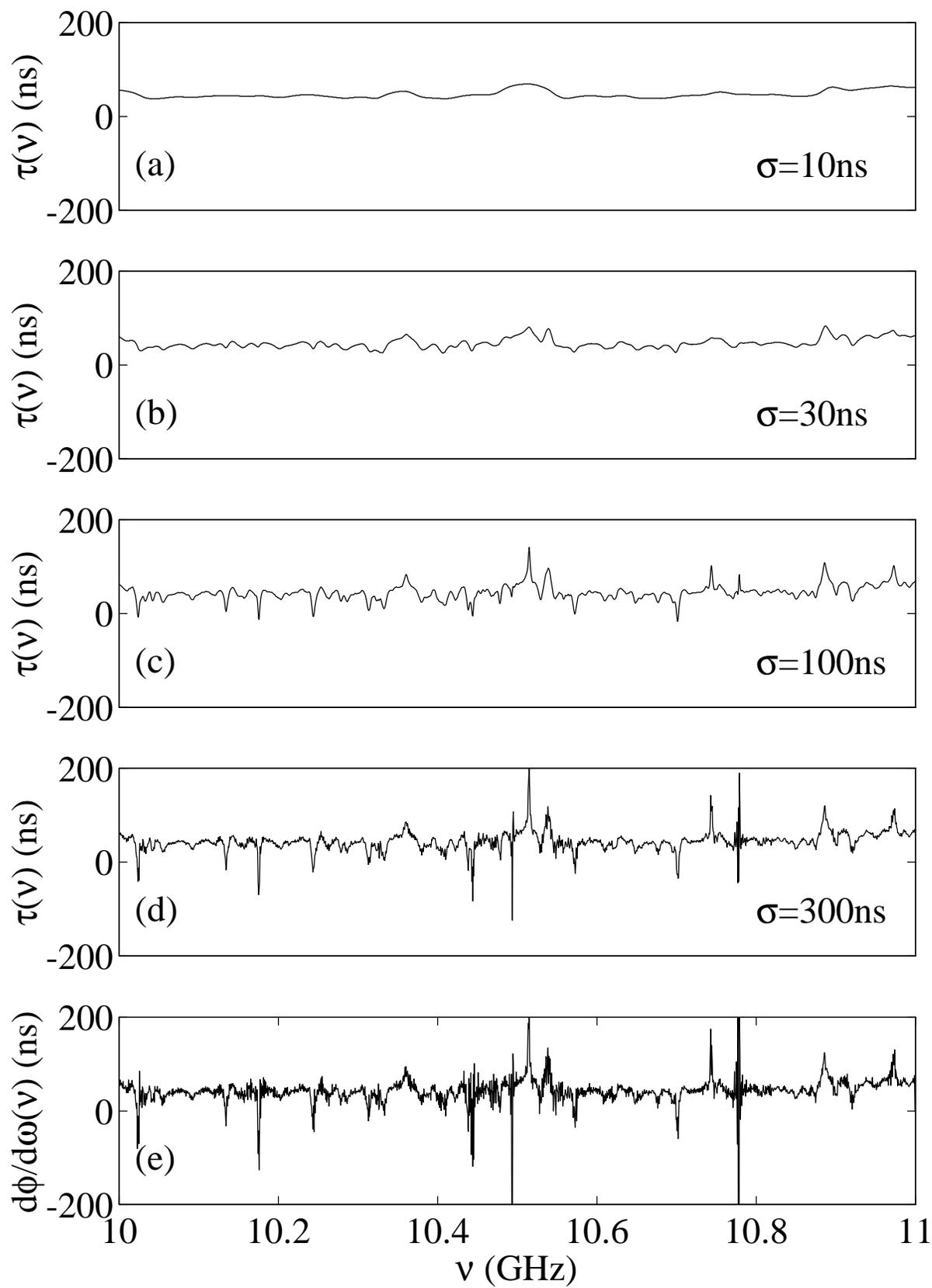
\end{document}